\documentclass[twocolumn,english,a4paper]{revtex4}
\usepackage[T1]{fontenc}
\usepackage[latin9]{inputenc}
\usepackage{geometry}
\usepackage{amsmath}
\usepackage{amssymb}
\usepackage{graphicx}

\makeatletter
\@ifundefined{textcolor}{}
{%
 \definecolor{BLACK}{gray}{0}
 \definecolor{WHITE}{gray}{1}
 \definecolor{RED}{rgb}{1,0,0}
 \definecolor{GREEN}{rgb}{0,1,0}
 \definecolor{BLUE}{rgb}{0,0,1}
 \definecolor{CYAN}{cmyk}{1,0,0,0}
 \definecolor{MAGENTA}{cmyk}{0,1,0,0}
 \definecolor{YELLOW}{cmyk}{0,0,1,0}
}

\usepackage{array}
\usepackage{multirow}
\usepackage{setspace}

\usepackage{multirow}
\usepackage{caption}
\usepackage{color}
\usepackage{epstopdf}
\usepackage{epsfig}
\usepackage[normalem]{ulem}
\usepackage{babel}
\usepackage{float}
\def \tb{\textcolor{black}}

\makeatother

\usepackage{babel}
\begin{document}
\title{Orientation dynamics of two-dimensional concavo-convex bodies}
\author{S. Ravichandran}
\email{sravichandran@iitb.ac.in}

\affiliation{Nordita, KTH Royal Institute of Technology and Stockholm University,
Stockholm, SE 10691}
\affiliation{Interdisciplinary Programme in Climate Studies, Indian Institute of Technology Bombay, Mumbai 400076}

\author{J. S. Wettlaufer}
\email{john.wettlaufer@su.se}

\affiliation{Yale University, New Haven, CT 06520-8109, USA}
\affiliation{Nordita, KTH Royal Institute of Technology and Stockholm University,
Stockholm, SE 10691}

\begin{abstract}
We study the orientation dynamics of two-dimensional concavo-convex
solid bodies more dense than the fluid through which they fall under gravity.
We show that the orientation dynamics of the body, quantified in terms
of the angle $\phi$ relative to the horizontal, undergoes a transcritical
bifurcation at a Reynolds number $Re_{c}^{(1)}$, and a subcritical
pitchfork bifurcation at a Reynolds number $Re_{c}^{(2)}$.
For $Re<Re_{c}^{(1)}$, the concave-downwards orientation of $\phi=0$
is unstable and bodies overturn into the $\phi=\pi$ orientation.
For $Re_{c}^{(1)}<Re<Re_{c}^{(2)}$, the falling body has two stable
equilibria at $\phi=0\text{ and }\phi=\pi$ for steady descent. For
$Re>Re_{c}^{(2)}$, the concave-downwards orientation of $\phi=0$
is again unstable, and bodies that start concave-downwards exhibit
overstable oscillations about the unstable fixed point, eventually
tumbling into the stable $\phi=\pi$ orientation. The $Re_{c}^{(2)}\approx15$
at which the subcritical pitchfork bifurcation occurs is distinct
from the $Re$ for the onset of vortex shedding, which causes the
$\phi=\pi$ equilibrium to also become unstable, with bodies fluttering
about $\phi=\pi$. The complex orientation dynamics of irregularly
shaped bodies evidenced here are relevant in a wide range of settings, from 
the tumbling of hydrometeors to settling of mollusk shells.  
\end{abstract}
\maketitle
%

% \section{Introduction}

Solid bodies falling under gravity through lighter fluids are ubiquitous, from tree leaves/seeds fluttering as they fall to the ground \citep[e.g.][]{Varshney2012}, to the dynamics of ammonia mushballs in the Jovian atmosphere \citep{Guillot2020}. Such objects are rarely spherical, typically exhibiting surfaces of non-uniform curvature which influence phenomena across the biological and physical sciences.  Biota, from plant to animal, are typically 
`concavo-convex' and, for example, irregularly shaped ice crystals can be crucial in determining the effective albedo, and hence the radiative balance, of the atmosphere \citep{Yang2015}.  However, the relatively few systematic studies of such concavo-convex bodies are at high Reynolds numbers. Allen \citep{ALLEN1984} studied the settling behavior of mollusk shells in sea water, finding that the shells universally settle with the concave side up. This behavior, along with the observation that sand grains may be trapped in the vortex bubble of the body's wake, is used to rationalize observations of concave-up shells on the sea bed.  However, because a concave-up shell on a substrate can re-orient in a shear flow, whereas a concave-down shell is more stable, of relevance to eggshell taphonomy \citep{Hayward2011}, post depositional re-orientation is an important mechanism in a variety of settings.

Concavo-convex bodies exhibit compelling dynamical phenomena across a wide range of flow conditions and Reynolds numbers, $Re=VL/\nu\gg1$, where $V$, $L$ and $\nu$ are body velocity, the body length \tb{scale} and the fluid viscosity respectively.  At low \tb{but nonzero $Re$, inertial torques act on symmetric large aspect ratio bodies such that they} settle with a horizontal long-axis, perpendicular to gravity \citep[e.g.,][]{Khayat1989}, whereas more complex bodies, such as dumbbells with different sized spheres, settle with a vertical long-axis and the large sphere at the bottom \citep{Candelier2016}, \tb{while rigid trumbbells settle with the `head' down (i.e. the centres of the spheres forming a `V' shape) \citep{ekiel_jezewska2009}, and flexible chains of spheres form a concave-upwards shape as they settle \citep[see, e.g.,][]{schlagberger_orientation_2005,bukowicki_different_2018}. }  At high $Re$, \tb{concavo-convex} bodies have been used as passive models to explain hovering flight.  For example, Childress and colleagues \citep{Childress2006,Weathers2010,Liu2012} showed that concave\emph{-downwards}($\wedge$) hollow pyramid-shaped bodies hover in oscillatory flows, whereas bodies that either start concave-upwards $\left(\vee\right)$, or are perturbed into this orientation, can no longer hover.  The asymmetry of the body, and the associated drag asymmetry, underlie the steady hovering observed in the $\wedge$ orientation. 

\tb{In this Letter,} we use well-resolved numerical simulations to study the orientation dynamics of falling two-dimensional concavo-convex solid bodies \tb{at $O(10)$ Reynolds numbers.}
Two horizontal orientations, $\phi=0$ (defined as the concave-downwards) and $\phi=\pi$, are possible for steady descent. At low $Re$, in analogy with \citep{ekiel_jezewska2009}, we expect that a body will settle concave-upwards. Whereas experiments \citep{ALLEN1984,Chan2020} suggest that only the concave-upward orientation is stable at moderate to large $Re$, we find that for a finite range of $Re$, the $\phi=0$ orientation is stable, indicating a stability boundary.  In particular, we find a transcritical bifurcation at $Re_{c}^{(1)}\approx2.5$, where the unstable $\phi=0$ orientation becomes a stable spiral, and a subcritical pitchfork bifurcation at $Re_{c}^{(2)}\approx15$, where the stable
spiral at $\phi=0$ becomes an unstable spiral.  Finally, we show that the critical Reynolds number $Re_{c}^{(2)}$ is distinct from and smaller than the Reynolds number at which vortices begin to be shed from the body. 

\begin{figure}
\includegraphics[width=0.8\columnwidth]{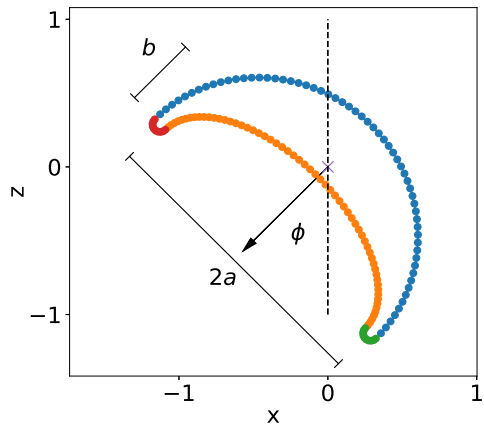}
\caption{\label{fig:schematic}The idealized concavo-convex body used here. The four elliptical sections are shown in different colors and correspond to (a) blue: a semi-ellipse with semi-major axis $a=1+\delta$ and unit semi-minor axis; (b) orange: a semi-ellipse with semi-major axis $a=1-\delta$ and semi-minor axis $b=0.5$; (c) red and green:
semi-circular sections with radius $\delta$. Here, $\delta=0.05$. The $\times$ denotes the location of the center of mass of the body.}
\end{figure}

Our body has density $\rho_{s}$, falls in the $-z$ direction through a fluid of density $\rho$, and is made up of elliptical or circular segments as shown in Fig. \ref{fig:schematic}, \tb{thereby ensuring that there are no corners. }  The perimeter of the body is discretized with marker points that, in addition to their normal orientations, are translated and rotated with the body velocity.  At each time step the body is ``reconstructed'' from these marker points and normals, as described by Engels et al. \citep{Engels2016}.  
The fluid velocity matches the local velocity of the body everywhere on the solid-fluid interface and thus the body is frictionally coupled to the flow field. 
Solving the associated equations of motion and boundary conditions challenge traditional numerical methods. An alternative is to model the
solid body as a porous medium with a vanishing porosity, which is achieved by introducing a term that forces the fluid velocity to relax to the local solid velocity exponentially, using a tunable penalization time constant $\eta$ \citep{Kevlahan2001}. This volume-penalization can be incorporated into most existing Navier-Stokes solvers. 

We nondimensionalize the governing equations with the semi-major axis $a$ as the length scale, $U_{b}=\left[\left(\rho_{s}/\rho-1\right)ag\right]^{1/2}$
as the velocity scale and the fluid density $\rho$ as the density scale.  Thus, the equations of motion for the velocity field $\boldsymbol{u} = (u,w)$, are
\begin{align}
\frac{D\boldsymbol{u}}{Dt} & =-\nabla p+\frac{1}{Re}\nabla^{2}\boldsymbol{u}+\chi\frac{\boldsymbol{v-u}}{\eta},\text{ and }\label{eq:momentum}\\
\nabla\cdot\boldsymbol{u} & =0,\label{eq:continuity}
\end{align}
where $Re=U_{b}a/\nu$ is the Reynolds number, \textbf{$\boldsymbol{v}\left(\boldsymbol{x},t\right)$} is the local velocity of the solid body, and the mask function is $\chi=1$
in the region of space occupied by the solid and vanishes elsewhere.
The coordinates of the body center of mass, $\boldsymbol{x}_{g}$, and the angle the major axis makes with the horizontal, $\phi$, obey
\begin{align}
m_{b}\frac{d^{2}\boldsymbol{x}_{g}}{dt^{2}} & =\boldsymbol{f}_{b}+ \tb{m_b}\boldsymbol{g}\label{eq:body_vel}\qquad\text{and}\\
I_{b}\frac{d^{2}\phi}{dt^{2}} & =\tau_{b},
\end{align}
where  the buoyancy-corrected mass and moment of inertia of the body are  \tb{$m_b= (\rho_s - \rho)\int d\mathbf{x} = (\rho_s - \rho) V$, and $I_b= (\rho_s - \rho) \int{\mathbf{(x-x_g)^2} d\mathbf{x}}$} respectively.  Here, $V$ is the volume per unit length of the body; the forces \tb{$\boldsymbol{f}_{b} = \rho \eta^{-1} \int \mathbf{(u-v)} d\mathbf{x}$
and torque $\tau_{b}  = \rho \eta^{-1} \int \mathbf{(x-x_g) \times (u-v)}$ } exerted by the fluid on the body are calculated as volume integrals \cite[see e.g.,][]{Kolomenskiy2009}. 

We solve Eqs \eqref{eq:momentum}--\eqref{eq:continuity} in two-dimensions using a Fourier-pseudospectral method as described in \citep{Ravichandran2017b} (validation studies are presented in the supplementary material).  We use an $x-z$ domain of $40\times80$ (or $40\times160$ for long-time simulations) and find that periodic boundary conditions do not affect the dynamics beyond this domain size. The $x-z$ domain is discretized with $2048\times4096$ (or $2048\times8192$) gridpoints, so that there are $>50$ points per unit length. We have verified that the results are insensitive at this grid resolution, and small variations thereof, with a volume penalization parameter of $\eta=10^{-3}$.
The concavo-convex bodies are released from rest with the major axis tilted at an initial angle $\phi\left(t=0\right)\equiv\phi_{0}$ relative to the horizontal as shown in Fig. \ref{fig:schematic}.  Thus, for a given $Re$, finite initial angles act as a perturbations to a fixed point, which we observe to grow or decay as described presently. \tb{Our results are qualitatively similar for variations in the aspect ratio of the body $b/a$. A detailed study of how the body shape parameters affect stability will be presented elsewhere.}

\begin{figure}
\begin{centering}
\includegraphics[width=0.45\columnwidth]{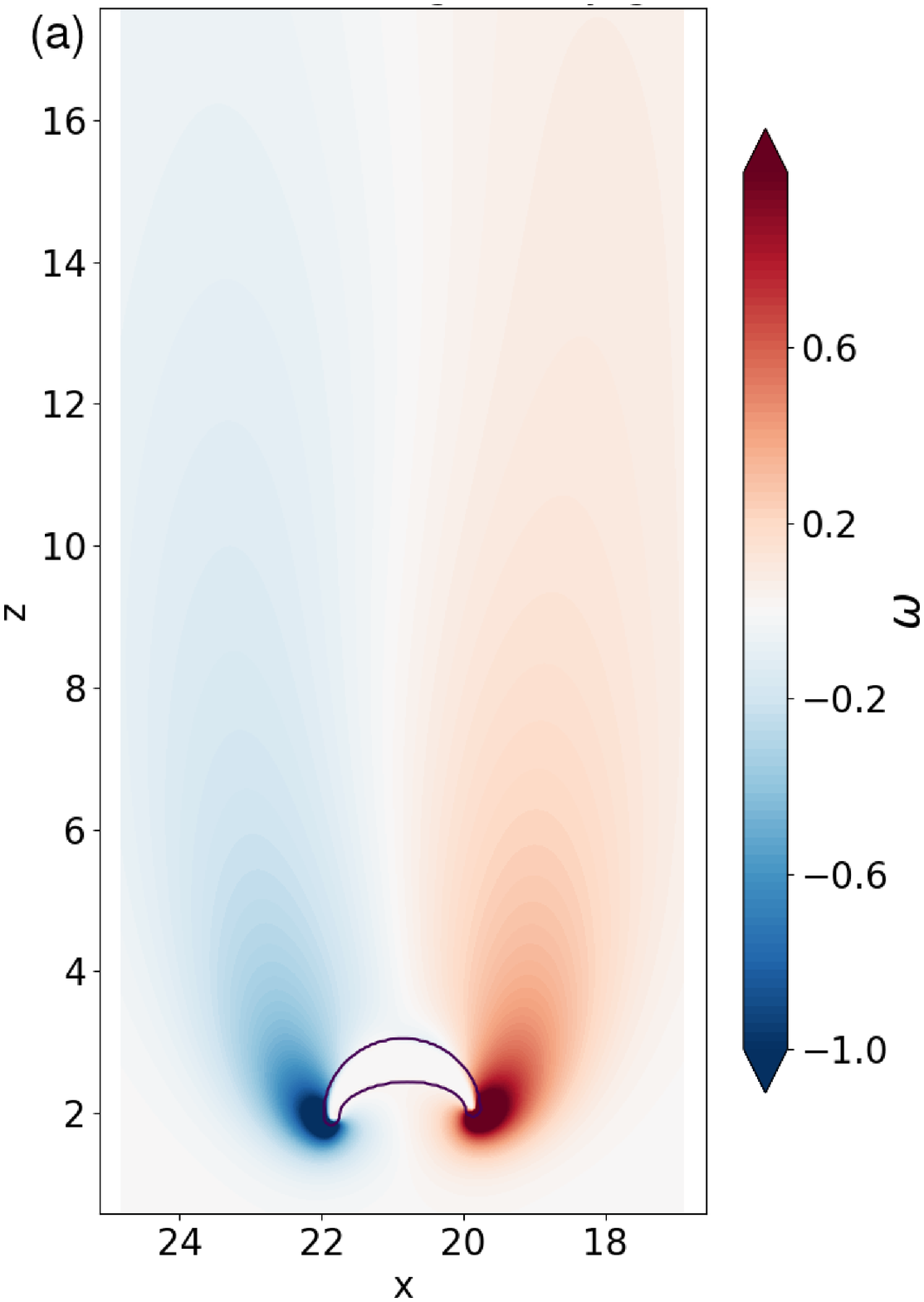}\includegraphics[width=0.45\columnwidth]{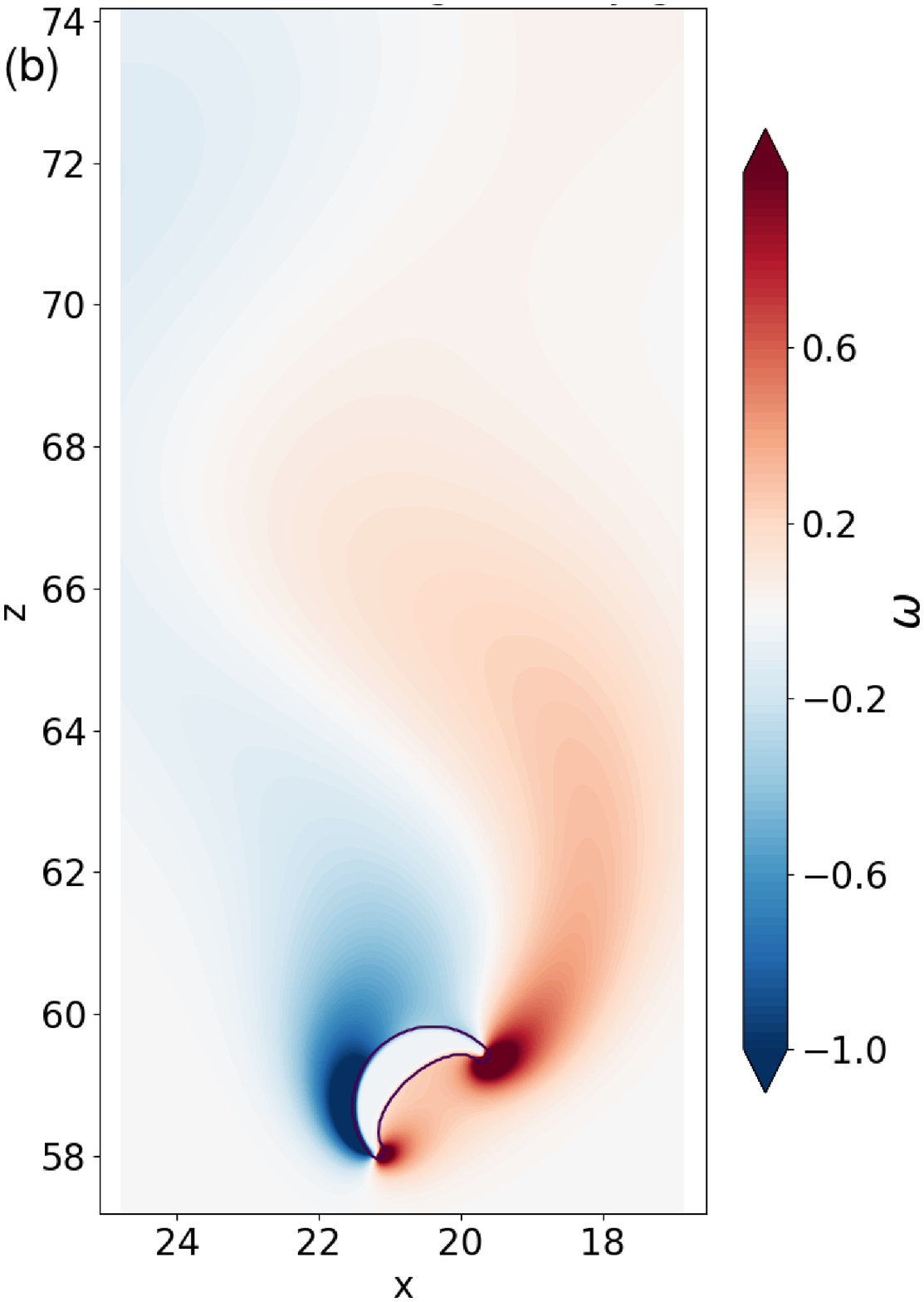}
\par\end{centering}
\begin{centering}
\includegraphics[width=0.45\columnwidth]{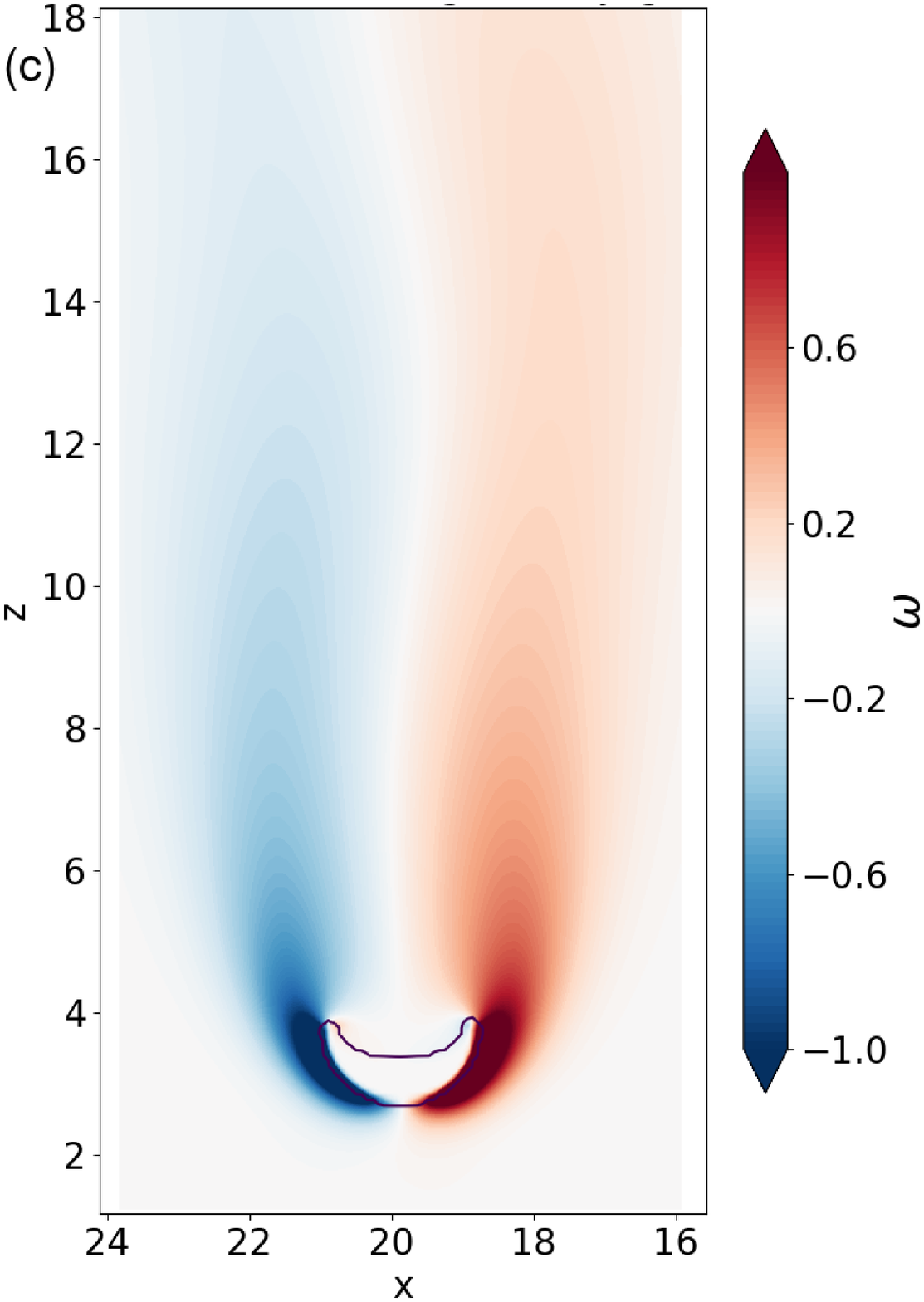}\includegraphics[width=0.45\columnwidth]{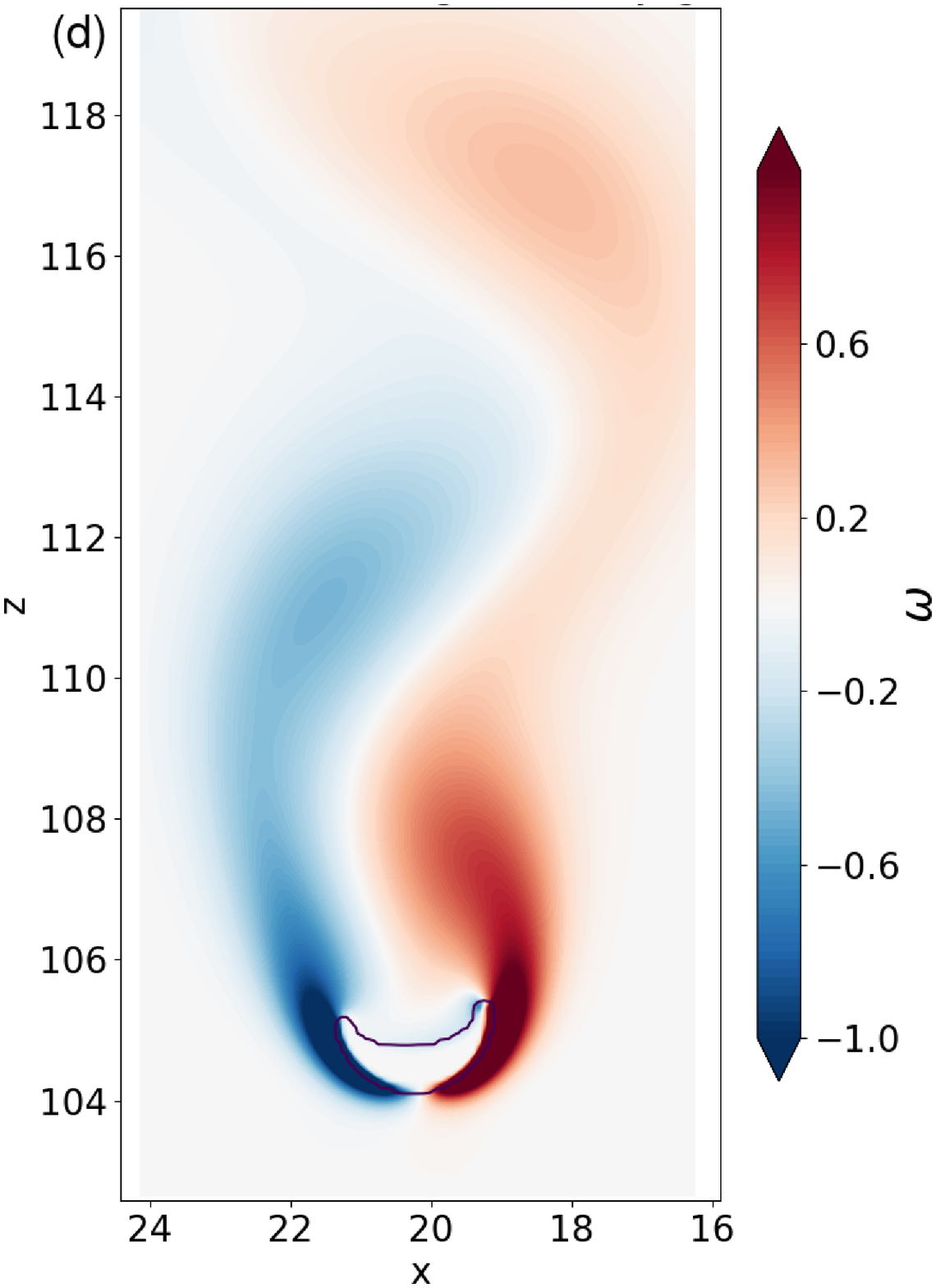}
\par\end{centering}
%\raggedleft{}
\caption{\label{fig:omg_chi_vs_xz} The vorticity, $\omega_y$,
shown in color and the outline of the concavo-convex body
shown as the solid black line. (a) For $Re=10<Re_{c}^{(2)}$, the
$\phi=0$ orientation is a stable fixed point. (b) For $Re=12$, the initial orientation
$\phi_{0}=\pi/12$ is outside the basin of attraction of the fixed
point at $\phi=0$. As the oscillations increase in amplitude they are reflected in the deviations of the
dividing streamline from the vertical,
and vortices are shed. (c) For $Re=17>Re_{c}^{(2)}$, the $\phi=0$
orientation is unstable and all initial orientations $\phi_{0}$ lead
to steady descent at $\phi=\pi$. Note, however, this $Re$ is smaller than
required for the onset of vortex shedding. (d) For $Re=25$, in the
concave-up orientation, an incipient Karman vortex street is seen
and the system oscillates with a finite amplitude.}
\end{figure}

In Fig. \ref{fig:omg_chi_vs_xz} we show contour plots of the $y$ component of the vorticity, 
%\begin{equation}
$\omega_y=\frac{\partial u}{\partial z}-\frac{\partial w}{\partial x}$, 
%\label{eq:omg_y}
%\end{equation}
around the body for $Re=10,12,17$ and $25$.  The following are our four key observations. 

\begin{figure}
\begin{centering}
\includegraphics[width=0.9\columnwidth]{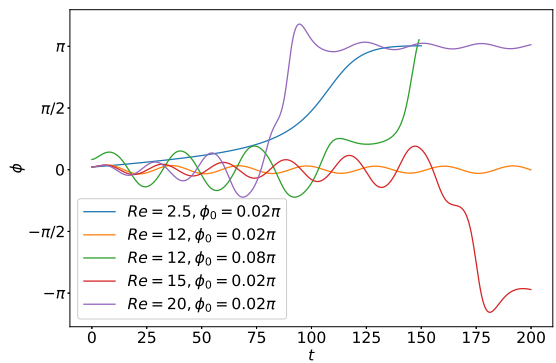}
\par\end{centering}
\raggedleft{}\caption{\label{fig:phi_vs_t_Re} The orientation evolution $\phi\left(t\right)$ of
the bodies (see Fig. \ref{fig:schematic}). The final orientation
is determined by the combination of the Reynolds number $Re$ and
the initial orientation $\phi_{0}$. For small Reynolds number, $Re=2.5<Re_{c}^{(1)}$,
$\phi$ increases monotonically with $t$ for all initial orientations
$\phi_{0}$, overshooting $\phi=\pi$ slightly before reaching $\phi=\pi$.
For $Re_{c}^{(1)}<Re=12<Re_{c}^{(2)}$ the trajectories are oscillatory;
for small $\phi_{0}=0.07,$ $\phi(t)$ decays to zero as the body
reaches a state of steady descent, while for $\phi_{0}=\pi/12$ the
body tumbles into the $\phi=\pi$ orientation before descending steadily.
For $Re>Re_{c}^{(2)}\approx15$, the body flips to the $\phi=\pi$
orientation for all $\phi_{0}$, and the time in flight before the
body tumbles decreases with increasing $Re$ and increasing $\phi_{0}$.}
\end{figure}

\begin{figure}
\includegraphics[width=0.95\columnwidth]{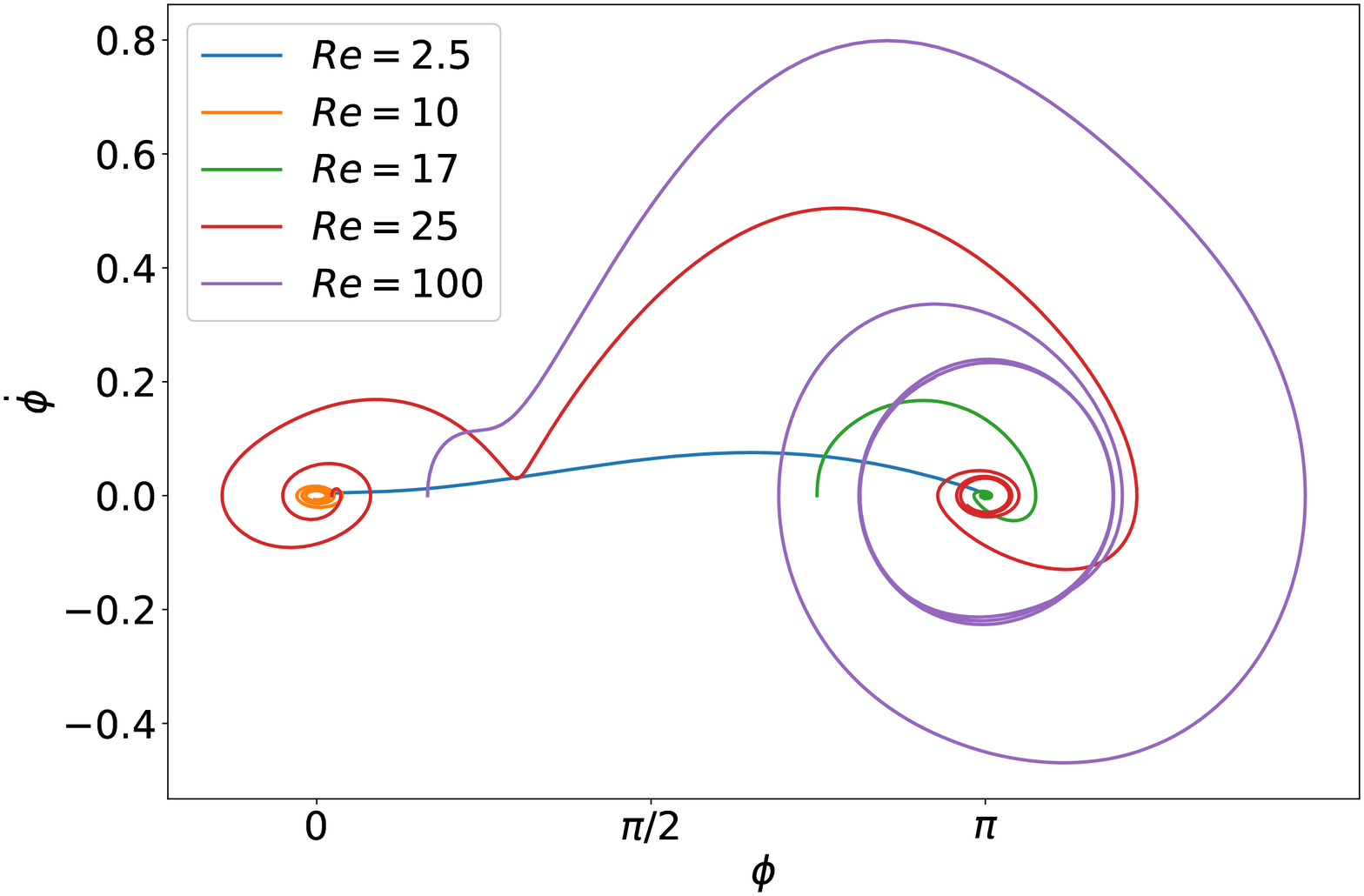}

\caption{\label{fig:phi_vs_phidot} Phase space plots of the angular velocity
$\dot{\phi}$ versus the angle $\phi$ for Reynolds
numbers $Re=2.5,10,17,25,100$. For $Re=2.5$, the orientation $\phi=0$
is unstable and the trajectory $\dot{\phi}\left(\phi\right)$ towards
$\phi=\pi$ is non-oscillatory. For $Re=10$, the orientation $\phi=0$ is
a stable spiral and the trajectory spirals inwards. For $Re=17$,
perturbations about $\phi=\pi$ decay to zero and the trajectory is
a decaying spiral around $\left(\phi=\pi,\dot{\phi}=0\right)$. For
$Re=25$, the system is overstable about the $\phi=0$ orientation,
and the trajectory spirals outwards, reaching the $\phi=\pi$ orientation
where it becomes a limit cycle of finite radius. A limit cycle of
larger radius is reached for $Re=100$. }
\end{figure}

(1): \underline{$Re<Re_{c}^{(1)}$}. The $\phi=0$ orientation is unstable, while $\phi=\pi$
is stable. Bodies with all initial orientations overturn into the
$\phi=\pi$ orientation. The trajectories $\dot{\phi}\left(\phi\right)$
\emph{do not} spiral outwards from  zero, smoothly converging to $\phi=\pi$ (see Figs. \ref{fig:phi_vs_t_Re} and \ref{fig:phi_vs_phidot}),
%but small fluctuations beyond $\phi=\pi$ are observed.

(2): \underline{$Re_{c}^{(1)}<Re<Re_{c}^{(2)}$}.  Orientations $\phi=0$ and $\phi=\pi$ are both \emph{stable spirals}, with $Re$ dependent basins of attraction. As shown in Fig. \ref{fig:phi_vs_t_Re}, the initial orientation, $\phi_{0}$, determines which equilibrium is attained.  In particular, the basin of attraction, which is the set of values $\phi_{0}$ that converge to the $\phi=0$ fixed point, shrinks to zero as $Re$ increases to approximately $15$.
For $\phi_{0}$ within the basin of attraction of $\phi=\pi$ but
concave-downwards (for example see $Re=12,\phi_{0}=\pi/12$ in Fig. \ref{fig:phi_vs_t_Re}),
the oscillations of the body are associated with the significant meandering
of the streamline separating the $\omega_y<0$ and $\omega_y>0$ regions
(see Fig. \ref{fig:omg_chi_vs_xz}b). 

%\vspace{-0.25cm}

(3): \underline{$Re_{c}^{(2)}<Re\lesssim20$}. The orientation $\phi=\pi$ is \emph{stable}, with the amplitude of oscillations in $\phi$
decaying with time. This is seen in Fig. \ref{fig:phi_vs_phidot}, showing that trajectories in the phase space $\left(\phi,\dot{\phi}\right)$
spiral inwards to the point $\phi=\pi,\dot{\phi}=0$. Figure \ref{fig:omg_chi_vs_xz}(c)
also shows that $Re=17$ is too small to initiate vortex shedding in the $\phi=\pi$ orientation which, since the convex outer boundary has semi-major and semi-minor axes $b=a=1$, should occur at $Re\approx20$. 

(4): \underline{$Re\gtrsim20$}. The point $\phi=\pi$ is weakly unstable, due to the periodic shedding of vortices behind the body as shown in Fig. \ref{fig:omg_chi_vs_xz}(d). The trajectories in $\left(\phi,\dot{\phi}\right)$ space reach limit cycles whose amplitudes increase with increasing $Re$, as seen in Fig. \ref{fig:phi_vs_phidot}. The periodic oscillations
about $\phi=\pi$ are consistent with the experimental observations of curved bodies by Chan et al. \citep{Chan2020}. The weak instability at $\phi=\pi$ is also reminiscent of the behavior of passively hovering bodies in oscillatory flow \citep[e.g.,][]{Childress2006} where the concave-downwards orientation is stable.

\begin{figure}
\includegraphics[width=0.8\columnwidth]{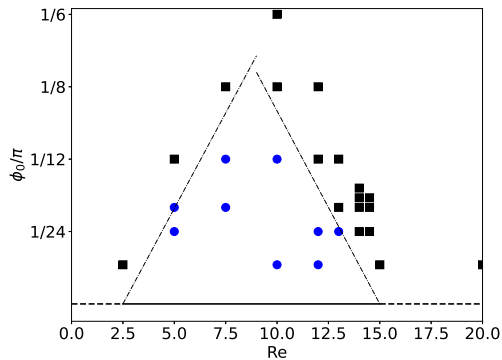}
\caption{\label{fig:bifurcation} Bifurcation diagram for $\left(\phi,\dot{\phi}\right)=\left(0,0\right)$
as the initial orientation, $\phi_{0}\protect\geq0$, and the Reynolds
number $Re$ are varied. Blue circles denote initial conditions that
are attracted to $\left(\phi,\dot{\phi}\right)=\left(0,0\right)$, while black squares
denote initial conditions that are attracted to $\left(\phi,\dot{\phi}\right)=\left(\pi,0\right)$.
The dot-dashed lines separating the circles from the squares are separatrices
drawn by hand. For $Re_{c}^{(1)}<Re<Re_{c}^{(2)}$, two stable fixed
points, $\phi=0$ and $\phi=\pi$ exist. At the subcritical pitchfork
bifurcation at $Re=Re_{c}^{(2)}$, the $\phi=0$ ($\phi=\pi$) fixed point becomes
unstable (remains stable). At a
higher $Re$ associated with vortex shedding, the $\phi=\pi$ fixed
point becomes unstable and a limit cycle in $\left(\phi,\dot{\phi}\right)$
appears (Fig. \ref{fig:phi_vs_phidot}).}
\end{figure}

Fig. \ref{fig:bifurcation} summarizes the points (1)-(4) as follows.  The point $\left(\phi,\dot{\phi}\right)=\left(0,0\right)$ undergoes a transcritical bifurcation at $Re\approx2.5$ and a subcritical pitchfork bifurcation at $Re\approx15$.  The point $\left(\phi,\dot{\phi}\right)=\left(\pi,0\right)$ undergoes a supercritical Hopf bifurcation at a higher $Re\gtrsim20$.

\begin{figure}
\includegraphics[width=\columnwidth]{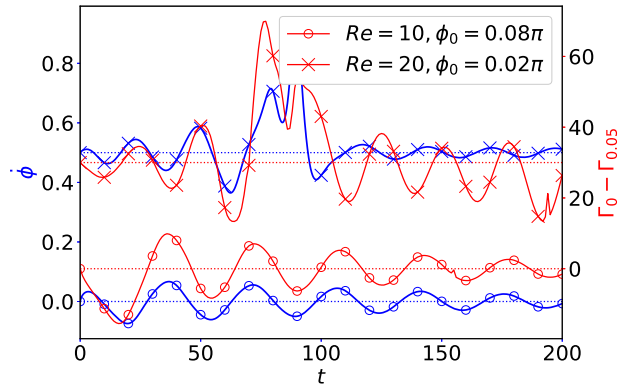}
\caption{\label{fig:phidot_circ}The angular velocity (blue, left axis) and the circulation $\Gamma_0 - \Gamma_{0.05}$ (red, right axis) for $Re=10$ and $Re=20$ with initial angles $\phi_0$ to the vertical. The latter curves are displaced vertically for clarity. The former (latter) initial condition is stable (unstable) in the concave-downwards orientation.}
\end{figure}

{The dynamics of the body are controlled by the torques exerted by the fluid upon it, which are related to the circulation $\Gamma_d$ on a contour at a distance $d$ normal to the surface, which includes the velocity boundary layer. \tb{The results are insensitive to the choice of $d$}. In Fig. \ref{fig:phidot_circ}, we plot {$\Gamma_0 - \Gamma_{0.05}$} for $Re=10$ and $Re=20$, where {$\Gamma_0=\Gamma_{d=0}$} is the circulation at the surface of the body and, by the Stokes theorem, is equal to the angular velocity of the body times its area. The initial orientation is close to the equilibrium $\phi=0$, and we see that $\Gamma_d$ decays with time for $Re=10$. For $Re=20$, the initial (concave-down) orientation is unstable, and $\Gamma_0 - \Gamma_d$ shows the oscillatory increase associated with the instability (see Fig. \ref{fig:phi_vs_t_Re}). We also find that whilst $\Gamma_0 - \Gamma_d$ and $\dot \phi$ are in phase for $Re=10$, $\Gamma_0 - \Gamma_d$ \emph{lags behind} $\dot \phi$ by a few flow time units. Thus, as the body reaches an extremity in its oscillation, the opposing torque continues to be nonzero, thus providing an impulse for the reverse motion and leading to an increase in the oscillation amplitude. We note that this oscillatory instability is not seen in lumped-mass approximations for the concavo-convex bodies as discussed in the Supplementary Material Sec. II.}

Our results show that over a finite range of $Re$, the $\phi=0$ fixed point becomes a stable spiral.  The $\phi=0$ orientation becomes a stable spiral
for $Re>Re_{c}^{(1)}$ due to the transient torque's discussed above.  For sufficiently small $Re$, a falling concavo-convex body may be expected to settle into the $\phi=\pi$ orientation because of inertial torques \citep{ekiel_jezewska2009,Candelier2016}. We explain the subcritical pitchfork bifurcation at $Re_{c}^{(2)}$ in analogy to the fluttering-to-tumbling transition seen in the dynamics of thin plates and disks, which have been more widely studied at higher $Re$.  
These studies find steady, oscillatory and chaotic dynamics as $Re$ is increased, wherein 
bodies of sufficiently large (nondimensional) moments of inertia exhibit tumbling \citep{Tanabe1994,Field1997,Belmonte1998,Mahadevan1999,Andersen2005,Kanso2014}.
Of particular note are the experiments of Belmonte et al., \citep{Belmonte1998}, with
two-dimensional thin flat plates of length $L$, thickness $d$, and width $w$ (in the third, homogeneous, dimension). The time scale for ``fluttering'', or side-to-side oscillation, is given by that of a buoyant pendulum $\tau_{p}=\left[\left(\rho_{s}/\rho-1\right)L/g\right]^{1/2}$,  assuming that the drag
on the plates is $F_{d}=\rho C_d V^{2}A$, with $V$ the terminal velocity, $A$ the cross sectional area, and $C_d$ a constant of order unity.  Thus $\tau_{v}=L/V$ is the timescale for vertical descent and hence when the body descends by a body length before it has completed a full oscillation, the plate tumbles
instead of fluttering, characterized by the Froude number, $Fr=\tau_{p}/\tau_{v}>Fr_{c}={O}\left(1\right)$.  Mahadevan et al., \citep{Mahadevan1999} show that this behavior may also be interpreted using the dimensional argument that balances the drag against gravity.  All other things being equal, narrower (in the third dimension) plates tumble end over end more readily than wider plates.   Belmonte et al. \citep{Belmonte1998} study a range of $Re$ from about $3 \times 10^{3} \text { to } 4 \times 10^{4}$ and note that their experiments are in agreement with previous observations that the motion is $Re$ independent, so long as $Re$ is sufficiently large.  Here, however, we examine $Re$ much smaller than did Belmonte et al. \citep{Belmonte1998}.  Thus, if we modify the drag law and assume that $C\sim Re^{-1}\sim\nu/\left(LV\right)$, then $V\sim\left(\rho_{s}/\rho-1\right)gLd/\nu$, leading to $Fr=\tau_{p}/\tau_{v}\sim Re^{1/2}$.  Therefore, whilst Belmonte et al., \citep{Belmonte1998} find a critical value of $Fr_{c}={O}\left(1\right)$ corresponding to a pendulum driven at resonance, their $Fr$ scaling shows that the fluttering is independent of viscosity.  This suggests that there must be a critical value for the Reynolds number {\em above which} the (steady) oscillation amplitude is large enough for the body to flip into the more stable concave-up orientation.

The smaller Reynolds numbers studied here are characteristic of systems with smaller length scales, such as water droplets or ice particles in clouds \cite{pruppacher1980} or in Saturn's rings \cite{goldreich1978, Esposito2010}, and a wide range of bio-particles settling in the ocean \citep{Guasto2012,Wheeler2019}.  Of particular relevance is the dependence of the stability of the $\phi=0,\pi$ equilibria as the dimensions of the body are altered. We note, for instance, that the concave-upward orientation is more resistant to tumbling than a flat plate. In organisms where passive stability in a certain orientation is beneficial, particular physiological dimensions or shapes may confer fitness, and thus be evolutionarily favored (see, e.g. \cite{chirat2021}). It would therefore be of interest to know if and how the stability boundary
$\left(Re_{c}^{(1)},Re_{c}^{(2)}\right)$ varies with the thickness of the body. Furthermore, the trapping of small particles in the body's
wake, hypothesized by Allen \citep{ALLEN1984}, suggests that particle segregation
by size may be achievable by simply dropping suitable shaped concavo-convex
bodies into a suspension. The trapping of small heavy particles in such flows is of considerable general interest in particle transport
(see, e.g. \citep{Ravichandran2014,Angilella2017}), thereby focusing attention on the role of particle shape, size and their distribution.  
Finally, it is clear from this work that extension to three dimensions, where a body may precess about
the vertical in addition to fluttering as described here, is of significant basic and practical interest. 
\vspace{-1.0cm}
\section*{Acknowledgements}
\vspace{-0.5cm}
Support from Swedish Research Council under grant no. 638-2013-9243 and computational resources from the Swedish
National Infrastructure for Computing (SNIC) under grants SNIC/2019-3-386, SNIC/2020-5-471, SNIC/2021-5-449 are gratefully acknowledged. Computations were
performed on Tetralith. We thank Bernhard Mehlig for an interesting discussions on the fluid mechanics of particles in flows and Lidya Tarhan on the general area of mollusk shells.  Nordita is partially supported by Nordforsk. 
\vspace{-0.5 cm}
% \bibliographystyle{unsrt}
% \bibliography{references}

\end{document}